\documentclass[journal]{IEEEtran}
\ifCLASSINFOpdf
\else
\fi

\usepackage{amssymb}
\usepackage{pifont,calc}
\usepackage{cite}

\usepackage{graphicx}
\usepackage[cmex10]{amsmath}
\usepackage{array}

\usepackage{cases} 
\usepackage{bm} 
\usepackage{subfigure}
\usepackage{mathrsfs}
\usepackage{url}
\usepackage[ruled]{algorithm2e}
\usepackage{arydshln}
\usepackage{multirow}
\usepackage{multicol}

\makeatletter

\newcommand{\Rmnum}[1]{\expandafter\@slowromancap\romannumeral #1@}
\makeatother

\begin{document}
%
\title{Complexity Comparison between Two Optimal-Ordered SIC  MIMO Detectors  Based on Matlab Simulations}
%
%
%

\author{Yanpeng~Wu and Hufei~Zhu
\thanks{Y. Wu is with the Department of Information Science and Engineering, Hunan First Normal University, Changsha 410205, China (e-mail: xjzxwyp@hnfnu.edu.cn)}
\thanks{H. Zhu is with the College of Computer Science and Software, Shenzhen University, Shenzhen 518060, China (e-mail:
zhuhufei@szu.edu.cn).}}

%
%

\markboth{Journal of \LaTeX\ Class Files,~Vol.~14, No.~8, August~2015}%
{Shell \MakeLowercase{\textit{et al.}}: Bare Demo of IEEEtran.cls for IEEE Journals}
%



\maketitle

\begin{abstract}
This paper firstly introduces our shared Matlab source code that  simulates the two
optimal-ordered SIC detectors proposed
in \cite{zhfvtc2008} and \cite{BLASTtransSP2015}.
Based on our shared Matlab code,
we compare the computational complexities between
the two detectors in \cite{zhfvtc2008} and \cite{BLASTtransSP2015}
by
theoretical complexity calculations and numerical experiments.
We carry out theoretical complexity calculations to
 obtain the worst-case complexities
 for the two detectors  in \cite{zhfvtc2008} and \cite{BLASTtransSP2015}.
 Then from the theoretical  worst-case complexities,
  we make the conclusion that
 the detector proposed in \cite{BLASTtransSP2015}
requires $9 N^2$ more floating-point operations (flops)
than the detector proposed in \cite{zhfvtc2008},
where $N$ is the number of transmit antennas.
Our numerical experiments also show that
the detector in \cite{BLASTtransSP2015}
requires  more worst-case and average complexities
than the detector in \cite{zhfvtc2008}.
\end{abstract}

\begin{IEEEkeywords}
MIMO, optimal ordered SIC
detectors, fast algorithms, complexity comparison.
\end{IEEEkeywords}

%
\IEEEpeerreviewmaketitle

\section{Introduction}

\IEEEPARstart{F}{or}
multiple
input multiple output (MIMO) systems,  two
optimal ordered successive interference cancellation (SIC)
detectors have been proposed in \cite{zhfvtc2008} and \cite{BLASTtransSP2015},
respectively.  In \cite{transSPcomments}, it has been shown that
both optimal-ordered SIC detectors proposed in \cite{zhfvtc2008} and \cite{BLASTtransSP2015}
 require the same ${\rm O}(MN^2+N^3)$ complexity,
where $N$ and $M$ are the numbers of transmit and receive antennas, respectively.
This paper firstly introduces our shared Matlab source code, which simulates the two detectors
in \cite{zhfvtc2008} and \cite{BLASTtransSP2015}.
 In our shared Matlab code,
 each statement
 requiring floating-point operations (flops)~\footnote{As in \cite{zhfvtc2008}, flops in this paper means real flops.}
 is followed by a statement to count its
 flops, while the statement to count the flops is obtained manually.
Then the execution of the Matlab code can generate the exact total
complexity (in the number of flops) for each detector.


Our shared Matlab code is utilized in this paper to compare the complexities of
the two detectors  in \cite{zhfvtc2008} and \cite{BLASTtransSP2015} by numerical experiments.
  On the other hand, based on the
above-mentioned statements to count the flops,
  we carry out
  theoretical complexity calculations to
 obtain the worst-case ${\rm O}(MN^2+N^3)$ and ${\rm O}(MN+N^2)$ complexities
 for the two detectors  in \cite{zhfvtc2008} and \cite{BLASTtransSP2015}.
Then we compare the theoretical complexities of
the two detectors  in \cite{zhfvtc2008} and \cite{BLASTtransSP2015},
to determine  one with a lower complexity.


 \section{Matlab Code and Complexity Calculations for the Detectors Proposed in \cite{zhfvtc2008} and \cite{BLASTtransSP2015}}

 In our shared Matlab code,
 each statement
 requiring flops
 is followed by a statement to count the required complex multiplication, complex additions,
 real multiplications, real additions, real divisions, real square root operations, and multiplications between a real number and a complex number,
 which are denoted as cm, ca, rm, ra, rdiv, rsqrt and rcm, respectively. We set
 cm=6, ca=2, rm=1, ra=1, rdiv=1, rsqrt=1 and rcm=2, to compute the number of
 flops.

 Based on the
above-mentioned statements to count the required complexities,
  we carry out
  theoretical complexity calculations to
 obtain the worst-case
  ${\rm O}(MN^2+N^3)$ and ${\rm O}(MN+N^2)$ complexities
 for the two detectors  in \cite{zhfvtc2008} and \cite{BLASTtransSP2015},
 in the numbers of complex multiplications, complex additions and flops.
 Notice that we do the best to convert the complexities into
 the numbers of complex multiplications and complex additions, and only the complexities that cannot be
 converted (into the numbers of complex multiplications and complex additions) are measured by the numbers of flops.

Though most statements in our shared Matlab code can be obtained directly from the relevant algorithms described in \cite{zhfvtc2008} and \cite{BLASTtransSP2015},
there are still
some statements in our shared Matlab code 
which
cannot be obtained directly from the relevant descriptions in \cite{zhfvtc2008} and \cite{BLASTtransSP2015}.
Then in this section, we will also explain the algorithms for
some statements in our shared Matlab code, which have not be described in detail in
\cite{zhfvtc2008} and \cite{BLASTtransSP2015}.

In what follows, the computational complexity of $j$ complex
multiplications and $k$ complex additions will be denoted as
 $\left\lceil j,k  \right\rceil$,
which is simplified
 to $\left\lceil j  \right\rceil$ if $j=k$.
Moreover, notice that
\begin{subnumcases}{\label{n2n3flops289189}}
 \sum\limits_{i = 1}^{N}i=\frac{N(N+1)}{2}=\frac{N^2}{2}+\frac{N}{2} &  \label{n2flops289189}\\
 \sum\limits_{i = 1}^{N}i^2=\frac{N(N+1)(2N+1)}{6}=\frac{N^3}{3}+\frac{N^2}{2}+\frac{N}{6} &  \label{n3flops289189}
\end{subnumcases}
will be utilized to compute the ${\rm O}(MN^2+N^3)$ and ${\rm O}(MN+N^2)$ complexities.

\subsection{Matlab Code and Complexity Calculations for the Detector Proposed in \cite{zhfvtc2008}}

\begin{table*}[!t]
\renewcommand{\arraystretch}{1.3}
\caption{The Worst-Case/Fixed ${\rm O}(MN^2+N^3)$ and ${\rm O}(MN+N^2)$ Complexities of  the Optimal-Ordered SIC Detector Proposed in \cite{zhfvtc2008}} \label{table_example} \centering
\begin{tabular}{c|c|c|}
\bfseries  Step/Sub-step   & \bfseries Complexity    & \bfseries    Complexity  \\
\bfseries  Number   & \bfseries for  the $n^{th}$ Iteration  & \bfseries  for the Step \\
\hline
 \multirow{2}{*}{{\bfseries N1-b}}  &  & $\sum\limits_{i = 1}^{N}\left\lceil(N-i+1)M,(N-i+1)(M-1)\right\rceil$  \\
     &  &  $\approx \left\lceil\frac{1}{2}N(N+1) M,\frac{1}{2}N(N+1)(M- 1)\right\rceil$  \\
\hdashline
 \multirow{2}{*}{{\bfseries N1-c}}  & $2n$ flops and
  $\left\lceil\frac{n}{2}\right\rceil+\sum\limits_{i = 1}^{n-1}\left\lceil i-1 \right\rceil +$  & $\sum\limits_{n = 1}^{N}2n \approx N^2$ flops and
    \\
         & $\sum\limits_{i = 1}^{n-1} \left\lceil n-i-1 \right\rceil \approx\left\lceil n^2-\frac{5}{2}n \right\rceil$   &  $\sum\limits_{n = 1}^{N}\left\lceil n^2-\frac{5}{2}n \right\rceil \approx \left\lceil\frac{1}{3}N^3+\frac{1}{2}N^2-\frac{5}{4}N^2 \right\rceil =\left\lceil\frac{1}{3}N^3-\frac{3}{4}N^2 \right\rceil$  \\
\hdashline
  \bfseries N1-d &  & $(MN)$  \\
\hdashline
\multirow{2}{*}{{\bfseries N2}}  & $\left\lceil \frac{{{N^2}}}{4}  \right\rceil$ when $n=N$ & $\left\lceil \frac{{{N^2}}}{4}  \right\rceil  + \sum\limits_{n = 1}^{N - 1} \left\lceil {\frac{n}{2}}  \right\rceil$    \\
    & $\left\lceil {\frac{n}{2}}  \right\rceil$ when $n<N$ & $\approx \left\lceil \frac{{{N^2}}}{2}  \right\rceil$  \\
\hdashline
    \multirow{3}{*}{{\bfseries N3}} & 22n flops (to compute Givens matrices),     &   $\sum\limits_{n = 1}^N {22n} +\sum\limits_{n = 1}^N {3n} \approx 11N^2 + \frac{3}{2} N^2$ flops  and  $\sum\limits_{n = 1}^N \left\lceil{\frac{3}{2}}n^2-{\frac{9}{2}}n, {\frac{1}{2}}n^2-{\frac{3}{2}}n \right\rceil$   \\
  &3n flops and  $\sum\limits_{i = 1}^{n-1} \left\lceil 3i-3,i-1  \right\rceil$   &  $\approx \left\lceil{\frac{1}{2}}N^3+{\frac{3}{4}}N^2-{\frac{9}{4}}N^2, {\frac{1}{6}}N^3+{\frac{1}{4}}N^2-{\frac{3}{4}}N^2 \right\rceil$   \\
    &$\approx \left\lceil{\frac{3}{2}}n^2-{\frac{9}{2}}n, {\frac{1}{2}}n^2-{\frac{3}{2}}n \right\rceil$    & $=\left\lceil{\frac{1}{2}}N^3-{\frac{3}{2}}N^2, {\frac{1}{6}}N^3-{\frac{1}{2}}N^2 \right\rceil$  \\
\hdashline
 \bfseries N4 & $\left\lceil n  \right\rceil$ & $\sum\limits_{n = 1}^N {\left\lceil n  \right\rceil}  \approx \left\lceil \frac{{{N^2}}}{2}  \right\rceil$  \\
\hdashline
  \bfseries N6 & $\left\lceil  n \right\rceil$ & $\sum\limits_{n = 1}^N {\left\lceil n  \right\rceil}  \approx \left\lceil  \frac{{{N^2}}}{2} \right\rceil$ \\
\hline
  \bfseries Total Complexity & \multicolumn{2}{c|}{ $\frac{27}{2}N^{2}$ flops and $\left\lceil\frac{1}{2}MN^2+\frac{5}{6}N^3+\frac{3}{2}MN-\frac{3}{4}N^{2},\frac{1}{2}MN^2+\frac{1}{2}N^3+\frac{3}{2}MN-\frac{1}{4}N^{2}\right\rceil$}   \\
\hline
  \bfseries Total Flops & \multicolumn{2}{c|}{$\frac{27}{2}N^{2}+6(\frac{1}{2}MN^2+\frac{5}{6}N^3+\frac{3}{2}MN-\frac{3}{4}N^{2})+2(\frac{1}{2}MN^2+\frac{1}{2}N^3+\frac{3}{2}MN-\frac{1}{4}N^{2})=4MN^2+6N^3+12MN+\frac{17}{2}N^2$}   \\
\hline
\end{tabular}
\end{table*}

The  optimal-ordered SIC detector proposed  in \cite{zhfvtc2008} consists of steps N1-N7, where step N1 includes
sub-steps N1-a, N1-b, N1-c and N1-d.   In the shared Matlab code, sub-step N1-a has been omitted for simplicity~\footnote{In \cite{zhfvtc2008},
the columns in the channel matrix $\bf{H}$ are permuted in sub-step N1-a according to  the optimal detection order of the
adjacent subcarrier if MIMO OFDM systems are utilized, while
in \cite{BLASTtransSP2015}, the columns in $\bf{H}$ are permuted in increasing order of
their norms, or permuted equivalently by the
sorted Cholesky factorization. In the shared Matlab code, neither detectors in \cite{zhfvtc2008} and \cite{BLASTtransSP2015}
permutes the columns in $\bf{H}$ for fair comparison. Moreover, the method to permute $\bf{H}$ in \cite{zhfvtc2008} can be applied in
\cite{BLASTtransSP2015}, and vice versa.}.
Based on the statements to count the complexities  in our Matlab code to implement the detector in \cite{zhfvtc2008},
we calculate the worst-case ${\rm O}(MN^2+N^3)$ and ${\rm O}(MN+N^2)$ complexities  for the steps/sub-steps of the detector in
\cite{zhfvtc2008}, and the corresponding results are given in Table \Rmnum{1}.
Among the steps/sub-steps listed in Table \Rmnum{1}, steps N2 and N3 will be further described in this subsection,  since some details about
steps N2 and N3 have not been covered in  \cite{zhfvtc2008}.

In step N3,
the permuted inverse  Cholesky factor ${\bf{F }}_{n}$
is block upper-triangularized
by a unitary transformation to obtain ${{\bf{F }}_{n-1} }$ for the next iteration, as shown in equation (9) of \cite{zhfvtc2008}, i.e.,
\begin{equation}\label{equ9transform}
{\bf{F }}_{n} {\bf{\Sigma }} = \left[ {\begin{array}{*{20}c}
   {{\bf{F }}_{n-1} } & {{\bf{u}}_{n - 1} }  \\
   {{\bf{0}}_{n - 1}^T } & {\lambda _n }  \\
\end{array}} \right],
\end{equation}
where ${\bf{\Sigma }}$ is a unitary transformation, ${\bf{u}}_{n - 1}$
 is an $(n-1)\times 1$ column vector, and $\lambda _n$
 is a scalar. In  \cite{zhfvtc2008}, the unitary transformation ${\bf{\Sigma }}$ in
 (\ref{equ9transform})  is performed by a sequence of Givens rotations.
  The complex
Givens rotation described in
   lines 7-10 of the right column on
\cite[p. 46]{zhfvtc2008}
can  be denoted as
  \begin{equation}\label{givensForComplexRZhuhufeiPaper}
\left[\begin{array}{ll}d & e\end{array}\right]\left[\begin{array}{rr}{c} & {s} \\ {-s^*} & {c}\end{array}\right]=\left[\begin{array}{ll}0 & r \end{array}\right],
 \end{equation}
where $c$,    $s$  and $r$ satisfy
\begin{subnumcases}{\label{FastGivensCSCSCSzhf}}
c = |e|/\sqrt{|e|^{2}+|d|^{2}} &  \label{FastGivensCCCzhf}\\
s = e  d^{*}/(|e| \sqrt{|e|^{2}+|d|^{2}})  &  \label{FastGivensSSSzhf} \\
r = e\sqrt{|e|^{2}+|d|^{2}}/|e|.  &  \label{FastGivensSSSzhfforRrr}
\end{subnumcases}

  In \cite{zhfvtc2008}, the fast complex Givens rotation in \cite{Complex_Givens_Jul14b} is utilized.
 Thus in the shared Matlab code,  the fast complex Givens rotation described by  Algorithm 3
 in \cite{Complex_Givens_Jul14b} is utilized, which computes the Givens matrix
$\left[\begin{array}{rr}{c} & {s} \\ {-s^*} & {c}\end{array}\right]$
 by 22 flops, i.e.,
 15 real multiplications, 5 real additions, 1 real division and 1 real square root operation.
 In the $n^{th}$ ($n=N,N-1,\cdots,1$) iteration,
  $n-1$ Givens rotation are required in the worst case,
 and then it requires $22(n-1)\approx 22n$ flops  to compute
 $n-1$ Givens matrices, as shown in line 10  of Table \Rmnum{1}.

\begin{table*}[!t]
\renewcommand{\arraystretch}{1.3}
\caption{The Worst-Case/Fixed ${\rm O}(MN^2+N^3)$ and ${\rm O}(MN+N^2)$ Complexities of  the Optimal-Ordered SIC Detector Proposed in \cite{BLASTtransSP2015}} \label{table_example2} \centering
\begin{tabular}{c|c|c|c|}
  &  \bfseries  Step   & \bfseries Complexity    & \bfseries  Complexity  \\
   & \bfseries  Number   & \bfseries for the $i^{th}$ Iteration  & \bfseries  for the Step \\
\hline
 \multirow{4}{*}{ \rotatebox{90}{Table \Rmnum{1}}} & 1 &  & $\left\lceil MN  \right\rceil$  \\
 \cdashline{2-4}
 &11  &  & $\left\lceil \frac{{{N^2}}}{2}  \right\rceil$ \\
  \cdashline{2-4}
 &13  & $\left\lceil  N-i \right\rceil$ & $\sum\limits_{i = 1}^N {\left\lceil N - i  \right\rceil}  \approx \left\lceil \frac{{{N^2}}}{2}  \right\rceil$ \\
  \cdashline{2-4}
 &15  & $\left\lceil N-i  \right\rceil$ & $\sum\limits_{i = 1}^N {\left\lceil N - i  \right\rceil}  \approx \left\lceil \frac{{{N^2}}}{2}  \right\rceil$ \\
\hline
 \multirow{9}{*}{ \rotatebox{90}{Table \Rmnum{2}}} & 2 for  &  &$\sum\limits_{j = 1}^{N}\left\lceil(N-j+1)M,(N-j+1)(M-1)\right\rceil\approx$ \\
 & ${\bf{\Phi }}$ &  & $\left\lceil\frac{1}{2}N(N+1) M,\frac{1}{2}N(N+1) (M- 1)\right\rceil$  for ${\bf{\Phi }} = {{\bf{H}}^H}{\bf{H}} + {\sigma ^2}{\bf{I}}$  \\
  \cdashline{2-4}
 & 2 for  &  &$\sum\limits_{j = 1}^{N}2(N-j)\approx N^2$ flops and $\sum\limits_{j = 1}^{N}\left\lceil(N-j)(j-1)+\frac{j}{2}\right\rceil\approx$ \\
 & ${\bf{R }}$ &  & $\left\lceil\frac{1}{6}N^3-\frac{1}{4}N^2\right\rceil$ for ${\bf{R }} = cholesky({\bf{\Phi }})$  \\
  \cdashline{2-4}
  & 3 & $\sum\limits_{j = 1}^{i-1}\left\lceil i-j-1  \right\rceil \approx \left\lceil \frac{1}{2}i^2-\frac{3}{2}i  \right\rceil$  & $\sum\limits_{i = 1}^{N} \left\lceil  \frac{1}{2}i^2-\frac{3}{2}i \right\rceil \approx \left\lceil  \frac{1}{6}N^3-\frac{1}{2}N^2 \right\rceil$   \\
  \cdashline{2-4}
              & 5 &  & $\left\lceil \frac{{{N^2}}}{4}  \right\rceil$  \\
  \cdashline{2-4}
  &12  & 32(N-i) flops  &  $\sum\limits_{i = 1}^N {32(N - i)}  \approx 16 N^2$ flops \\
   \cdashline{2-4}
   & \multirow{3}{*}{13} &3$(N-i)$ flops and
   &  $\sum\limits_{i = 1}^N {3(N-i)} \approx \frac{3}{2} N^2$ flops  and    \\
   &  &$\sum\limits_{j = 1}^{N-i}\left\lceil  3j-3,j-1 \right\rceil \approx$    & $\sum\limits_{i = 1}^N \left\lceil{\frac{3}{2}}(N-i)^2-{\frac{9}{2}}(N-i), {\frac{1}{2}}(N-i)^2-{\frac{3}{2}}(N-i) \right\rceil\approx$ \\
   &  &$\left\lceil{\frac{3}{2}}(N-i)^2-{\frac{9}{2}}(N-i), {\frac{1}{2}}(N-i)^2-{\frac{3}{2}}(N-i) \right\rceil$    & $\left\lceil{\frac{1}{2}}N^3-{\frac{3}{2}}N^2, {\frac{1}{6}}N^3-{\frac{1}{2}}N^2 \right\rceil$   \\
  \cdashline{2-4}
 &18  & $\left\lceil {\frac{N-i}{2}}  \right\rceil$ & $\sum\limits_{i = 1}^N {\left\lceil \frac{N-i}{2}  \right\rceil}  \approx \left\lceil   \frac{{{N^2}}}{4} \right\rceil$  \\
\hline
 \multicolumn{2}{c|}{{\bfseries{Total Complexity}   }}  & \multicolumn{2}{c|}{$\frac{37}{2}N^{2}$ flops and $\left\lceil \frac{1}{2}MN^2+\frac{5}{6}N^3+\frac{3}{2}MN-\frac{1}{4}N^{2}, \frac{1}{2}MN^2+\frac{1}{2}N^3+\frac{3}{2}MN+\frac{1}{4}N^{2}  \right\rceil$}      \\
\hline
 \multicolumn{2}{c|}{{\bfseries{Total Flops}   }}  & \multicolumn{2}{c|}{$\frac{37}{2}N^{2}+6(\frac{1}{2}MN^2+\frac{5}{6}N^3+\frac{3}{2}MN-\frac{1}{4}N^{2})+2(\frac{1}{2}MN^2+\frac{1}{2}N^3+\frac{3}{2}MN+\frac{1}{4}N^{2})=4MN^2+6N^3+12MN+\frac{35}{2}N^2$}      \\
\hline
\end{tabular}
\end{table*}

 In the $n^{th}$ ($n<N$) iteration of the iterative detection phase,
 step N2 obtains the squared length of all the $n-1$ rows in  ${{\bf{F }}_{n-1}}$ by
 \begin{equation}\label{LengthModify}
 {\left| {{\bf{F }}_{n-1} }(j,:) \right|^2}={\left| {{\bf{F }}_{n}}(j,:) \right|^2 -\left| {{\bf{u}}_{n - 1} }(j) \right|^2},
 \end{equation}
 where~\footnote{As in \cite{BLASTtransSP2015}, the
MATLAB standard is followed in this paper to describe  algorithms.}
$j=1,2,\cdots,n-1$, and
   ${\left| \bullet \right|^2}$ denotes the squared length of a vector or  the squared absolute value  of a number.
   In  Appendix C, we will show that (\ref{LengthModify}) was implied in \cite{OriginalSqrtBlast}, and can be regarded as
a specific instance
 of equation (A-34) on
\cite[p. 120]{PhdThesis}~\footnote{Please notice that both \cite{OriginalSqrtBlast}
and
\cite{PhdThesis}
 were published before \cite{BLASTtransSP2015}.}.



%

From (\ref{LengthModify}), it can be seen that in the $n^{th}$ ($n<N$) iteration of the iterative detection phase,
 step N2 obtains the squared length of all the $n-1$ rows in  ${{\bf{F }}_{n-1}}$
 by $2(n-1)$ real multiplications and
 $2(n-1)$ real additions, which are equivalent to
  $(n-1)/2 \approx n/2$ complex multiplications and
 the same number of
 complex additions~\footnote{A complex multiplication includes 4 real multiplications and 2 real additions, while a complex addition includes 2 real additions.}, as shown in line 9 of Table \Rmnum{1}.
On the other hand,
 in the first  iteration with $n=N$,
 it requires about $N^2/4$ complex multiplications and
 the same number of complex additions to compute  the initial squared length of all
 the $N$ rows in the triangular ${{\bf{F }}_{N}}$, as shown in line 8 of Table \Rmnum{1}.

 Table \Rmnum{1} also gives the total number of worst-case
  ${\rm O}(MN^2+N^3)$ and ${\rm O}(MN+N^2)$ flops required by the detector in \cite{zhfvtc2008},
 which is
 \begin{equation}\label{totalFlopstrans2011}
4MN^2+6N^3+12MN+\frac{17}{2}N^2.
 \end{equation}

 \subsection{Matlab Code and Complexity Calculations for the Detector Proposed in  \cite{BLASTtransSP2015}}

 In \cite{BLASTtransSP2015},
the proposed optimal-ordered SIC detector consists of steps 1-3 and 10-19
in  Table \Rmnum{1}, and the details of step 2 in  Table \Rmnum{1} are described in
Table  \Rmnum{2}.
 Based on the statements to count the complexities  in our Matlab code to implement the detector in \cite{BLASTtransSP2015},
we compute the worst-case ${\rm O}(MN^2+N^3)$ and ${\rm O}(MN+N^2)$ complexities  for the steps of the detector in
\cite{BLASTtransSP2015}, and give the corresponding results in Table \Rmnum{2}.
   Table \Rmnum{2} includes the complexities of
   steps 1, 11, 13 and 15  in  Table \Rmnum{1} of \cite{BLASTtransSP2015},
 and  the complexities of  steps 2, 3, 5, 12, 13 and 18 in  Table \Rmnum{2}
   of \cite{BLASTtransSP2015}, since the complexities of those steps are
   ${\rm O}(MN+N^2)$ or ${\rm O}(MN^2+N^3)$.
   Among the steps listed in Table \Rmnum{2}, steps 2, 3, 12 and 13
   in  Table \Rmnum{2}
   of \cite{BLASTtransSP2015}
    will be further described in this subsection,  since some details about
those steps  have not been covered in  \cite{BLASTtransSP2015}.

 In the left column on \cite[p. 4630]{BLASTtransSP2015},  \cite{MatrixComputationBook}
 has been cited to describe the details about the implementation of
 the Cholesky factorization and the back-substitution  (to compute the inverse of the Cholesky factor).
 Then in our shared Matlab code,
 Algorithm 4.2.1 (GaxpyCholesky) on
 \cite[p. 164]{MatrixComputationBook} is utilized to implement the
 Cholesky factorization for step 2 in  Table \Rmnum{2}
   of \cite{BLASTtransSP2015}, and Algorithm 3.1.2 (Row-Oriented Back Substitution) on  \cite[p. 107]{MatrixComputationBook}
 is utilized to implement the back-substitution for step 3 in  Table \Rmnum{2}
   of \cite{BLASTtransSP2015}.

The Givens rotation in steps 12 and 13 in  Table \Rmnum{2} of \cite{BLASTtransSP2015}
is obtained by exchanging the columns of
the conventional
Givens rotation in \cite{MatrixComputationBook},
as mentioned in the third paragraph of the left column
 on \cite[p. 4628]{BLASTtransSP2015}.
Then  \cite{MatrixComputationBook} was cited again
 in lines 3-5 of the right column
on   \cite[p. 4630]{BLASTtransSP2015},
   to
claim
that  in step 13 in  Table \Rmnum{2} of \cite{BLASTtransSP2015},
 a conventional Givens rotation on a $(j+1)\times 2$ matrix  with complex entries
requires a complexity of 
\begin{equation}\label{flopsOfGivenswrong}
 \left\lceil  2j+2  \right\rceil.
 \end{equation}

 However,
  we can not find any description about the complex
 Givens rotation in \cite{MatrixComputationBook}, where only the real Givens rotation is introduced.
 Obviously the real  Givens rotation
 in \cite{MatrixComputationBook} can not be utilized to rotate the complex matrices in
 \cite{BLASTtransSP2015}. Accordingly, we refer to
 the next edition of the book \cite{MatrixComputationBook}, i.e., \cite{MatrixComputationBook4th},
and  follow
 the complex Givens rotation~\cite[equation (5.1.12) on p. 244]{MatrixComputationBook4th}
  \begin{equation}\label{givensForComplexROriginal}
\left[\begin{array}{rr}{c} & {s} \\ {-s^*} & {c}\end{array}\right]^{H}\left[\begin{array}{l}{u} \\ {v}\end{array}\right]=\left[\begin{array}{l}{
r} \\ {0}\end{array}\right].
 \end{equation}
In (\ref{givensForComplexROriginal}),
 $ ( \bullet )^* $ represents conjugate, and the real $c$ and the complex  $s$  satisfy
\begin{subnumcases}{\label{FastGivensCSCSCS}}
c = \cos (\theta ) &  \label{FastGivensCCC}\\
s = \sin (\theta ){e^{i\phi }}.  &  \label{FastGivensSSS}
\end{subnumcases}

 In \cite{transSPcomments},
 \cite{zhfvtc2008} has been cited to claim that
 the complexity of
(\ref{flopsOfGivenswrong})
should be revised
into the complexity of
\begin{equation}\label{flopsOfGivens1928}
\left\lceil  3j+3,  j+1  \right\rceil.
\end{equation}
In Appendix A,
we will
verify the complexity of (\ref{flopsOfGivens1928}).
Our shared Matlab code computes the Givens matrix 
 $\left[\begin{array}{rr}{c} & {s} \\ {-s^*} & {c}\end{array}\right]$
 and the corresponding result $r$ in (\ref{givensForComplexROriginal})
 by the conventional Givens rotation algorithm in \cite{MatrixComputationBook4th} ,
 which will be introduced in Appendix B. Appendix B also gives
 the conclusion that
 the conventional Givens rotation algorithm in \cite{MatrixComputationBook4th}
 computes $c$, $s$ and $r$ in (\ref{givensForComplexROriginal}) by
  20 real multiplications, 5 real additions, 4 real division and 3 real square root operation, which can be counted
  as
  32 flops.

Table \Rmnum{2} also gives the total number of worst-case
 ${\rm O}(MN^2+N^3)$ and ${\rm O}(MN+N^2)$ flops required by the detector in \cite{BLASTtransSP2015},
 which is
 \begin{equation}\label{totalFlopstrans2015}
4MN^2+6N^3+12MN+\frac{35}{2}N^2.
 \end{equation}

\section{Experiment and Discussion}

By comparing (\ref{totalFlopstrans2011})
and
(\ref{totalFlopstrans2015}),
we  can conclude that both the optimal-ordered SIC detectors proposed in \cite{zhfvtc2008}
and \cite{BLASTtransSP2015}  require the same dominant complexity, i.e.,
the  ${\rm O}(MN^2+N^3)$ complexity of
\begin{equation}\label{totalFlopsdominant}
4MN^2+6N^3
 \end{equation}
 flops. On the other hand,
it can be seen from (\ref{totalFlopstrans2011})
and
(\ref{totalFlopstrans2015}) that
the detector proposed in \cite{BLASTtransSP2015}
requires $9 N^2$ more flops
than the detector proposed in \cite{zhfvtc2008}.
 The above-mentioned
$9 N^2$ flops include $5 N^2$ flops caused by
the factor that the computation of the Givens matrices in
\cite{BLASTtransSP2015} requires
$16 N^2 - 11 N^2=5 N^2$ more flops than the corresponding computation
in  \cite{zhfvtc2008}, as can be seen
from Table \Rmnum{1} and Table \Rmnum{2}.
In the next paragraph, we will show that
above-mentioned
$9 N^2$ more flops also include $4 N^2$ flops to compute the initial
 $\underline{\mathbf{y}}=\left(\underline{\mathbf{R}}^{-1}\right)^{H} \underline{\mathbf{t}}$
 in
step 11 of Table \Rmnum{1} in \cite{BLASTtransSP2015}.


In this paragraph, let us compare the steps for the iterative detection phase in \cite{zhfvtc2008} and \cite{BLASTtransSP2015}.
  As step N4 in \cite{zhfvtc2008},
step 13 of Table \Rmnum{1} in \cite{BLASTtransSP2015} also forms the estimate and requires the same ${\rm O}(MN+N^2)$ complexity,
and as step N6 in \cite{zhfvtc2008},
step 15 of Table \Rmnum{1} in \cite{BLASTtransSP2015} also
cancels the effect of the detected signal
 and requires the same ${\rm O}(MN+N^2)$ complexity.
 However, step 11 of Table \Rmnum{1} in \cite{BLASTtransSP2015}, which computes the initial
 $\underline{\mathbf{y}}=\left(\underline{\mathbf{R}}^{-1}\right)^{H} \underline{\mathbf{t}}$ to
 cancel the effect of the detected signal in $\underline{\mathbf{y}}$ in the iterative detection phase,
 is not required in \cite{zhfvtc2008}, since instead of
 $\underline{\mathbf{y}}$ in \cite{BLASTtransSP2015},
 the
 detector in \cite{zhfvtc2008} cancels  the effect of the detected signal in $\underline{\mathbf{t}}$ (i.e., $\mathbf{z}_{m}$ in \cite{zhfvtc2008}).
Then  with respect to the detector in \cite{zhfvtc2008},
the detector in \cite{BLASTtransSP2015} requires the extra complexity of
 $\left\lceil \frac{{{N^2}}}{2}  \right\rceil$ (for step 11 of Table \Rmnum{1} in \cite{BLASTtransSP2015}) in  the iterative detection phase,
 which is equal to $ \frac{{{N^2}}}{2} \times (6+2)=4 N^2$ flops.



\begin{figure}[htbp]
\centering
\includegraphics[width=0.5\textwidth]{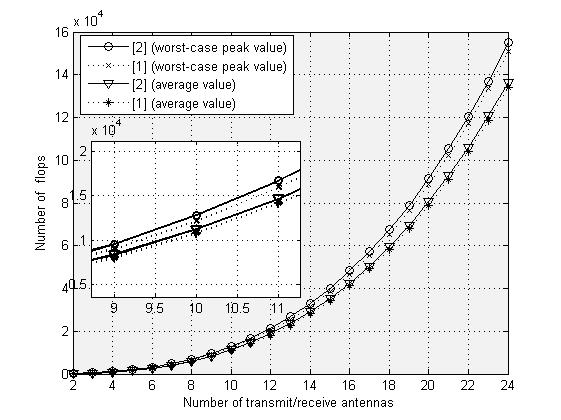}
\caption{Comparison of the worst-case and average complexities between
the two optimal-ordered SIC detectors proposed in
 \cite{zhfvtc2008} and  \cite{BLASTtransSP2015}.}
\end{figure}

Assume $N=M$. For different number of transmit/receive antennas, we
apply our shared Matlab code
to count the worst-case and average flops of
the optimal-ordered SIC detectors proposed in
 \cite{zhfvtc2008} and \cite{BLASTtransSP2015}. The results~\footnote{The  numbers of flops in Fig. 1 are usually less than  those in Fig. 1 of \cite{zhfvtc2008}.
This  can be explained by the fact that
the flops of
each statement in our shared Matlab code are counted
by hand  to give the exact number for Fig. 1 in this paper,
 while   the flops were counted by  the MATLAB 5.3 built-in function ``flops"
 to give a rough estimate for Fig. 1 in \cite{zhfvtc2008}.} are shown in Fig. 1.
As in \cite{BLASTtransSP2015}
and \cite{zhfvtc2008},
the maximum number of Givens rotations are assumed to count the worst-case flops.
To count the average flops, we simulate 10000  random
channel matrices $\bf{H}$, and  for fair comparison, we do not permute the columns of $\bf{H}$.
From Fig. 1, it can be seen that the detector proposed in \cite{BLASTtransSP2015}
requires  more worst-case and average flops
than the detector proposed in \cite{zhfvtc2008}, which
is  consistent with the theoretical flops calculation.

Fig. 2 shows the numbers of worst-case flops obtained by our shared Matlab code
and those computed by (\ref{totalFlopstrans2011}),
(\ref{totalFlopstrans2015}) and (\ref{totalFlopsdominant}).
From Fig. 2, it can be seen that
the numbers of worst-case flops obtained by our shared Matlab code
are very close to the theoretical numbers of
worst-case
  ${\rm O}(MN^2+N^3)$ and ${\rm O}(MN+N^2)$ flops computed by
  (\ref{totalFlopstrans2011})
  and
(\ref{totalFlopstrans2015}),
and are clearly larger than
the theoretical numbers of
worst-case
  ${\rm O}(MN^2+N^3)$ flops  computed by (\ref{totalFlopsdominant}).

%
%

\begin{figure}[!t]
\centering
\includegraphics[width=0.5\textwidth]{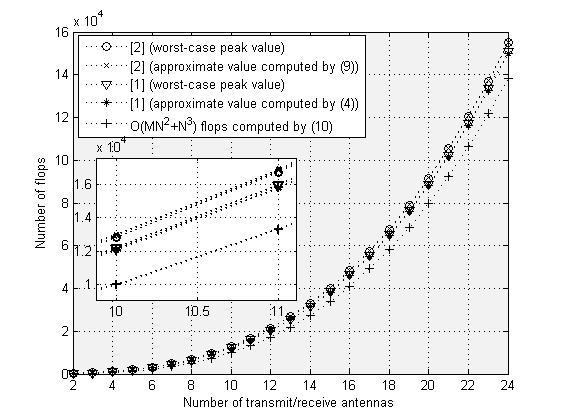}
\caption{Comparison of the worst-case complexities obtained by Matlab simulations,
 the theoretical  ${\rm O}(MN^2+N^3)$ and ${\rm O}(MN+N^2)$ worst-case complexities
 computed by (\ref{totalFlopstrans2011})
 and
(\ref{totalFlopstrans2015}), and the theoretical  ${\rm O}(MN^2+N^3)$ worst-case complexity
 computed by (\ref{totalFlopsdominant}).}
\label{fig_sim222}
\end{figure}

 \section{Conclusion}

In this paper,  we introduce our shared Matlab source code that  simulates the two
optimal-ordered SIC detectors proposed
in \cite{zhfvtc2008} and \cite{BLASTtransSP2015}.
We also explain some algorithms utilized in our shared Matlab code, which have not be described in detail in
\cite{zhfvtc2008} and \cite{BLASTtransSP2015}.

Based on our shared Matlab code,
we compare the computational complexities between
the two detectors in \cite{zhfvtc2008} and \cite{BLASTtransSP2015}
by theoretical complexity calculations and numerical experiments.
  We carry out theoretical complexity calculations to
 obtain the worst-case ${\rm O}(MN^2+N^3)$ and ${\rm O}(MN+N^2)$ complexities
 for the two detectors  in \cite{zhfvtc2008} and \cite{BLASTtransSP2015},
 from which we make the conclusion that
 the detector proposed in \cite{BLASTtransSP2015}
requires $9 N^2$ more flops
than the detector proposed in \cite{zhfvtc2008}.
Our
numerical experiments
 show that
 the detector in \cite{BLASTtransSP2015}
requires  more worst-case and average flops
than the detector in \cite{zhfvtc2008},
and the numbers of worst-case flops obtained by our shared Matlab code
are very close to the theoretical numbers of
worst-case  ${\rm O}(MN^2+N^3)$ and ${\rm O}(MN+N^2)$ flops.

\appendices

\section{To Verify the Complexity of (\ref{flopsOfGivens1928}) for a Complex Givens Rotation on a $(j+1)\times 2$ Matrix}

In this appendix,  we will utilize the fact that a complex multiplication is equivalent to $4$ real multiplications and $2$ real additions,
and a complex addition is equivalent to $2$ real additions.
We can
 write the left side of (\ref{givensForComplexROriginal})
as
 \begin{equation}\label{Givens1}
\left[\begin{array}{rr}{c} & {s} \\ {-s^*} & {c}\end{array}\right]^{H}\left[\begin{array}{l}{u} \\ {v}\end{array}\right]
=\left[\begin{array}{l}{{{ c u- sv}}} \\ {{s^*u+ cv}}\end{array}\right].
 \end{equation}
Since $c$ is real, the complexity of (\ref{Givens1}) should be less than 4 complex multiplications and 2 complex additions.
In (\ref{Givens1}), it requires  $2$ complex multiplications to compute $s\times v$ and $s^* \times u$,
$4$ real multiplications to compute $c\times u$ and $c \times v$,
and $4$ real additions to compute ${{c u- sv}}$ and ${{s^*u+ cv}}$.
Then the total complexity of (\ref{Givens1}) is equivalent to $2 \times 4 + 4=12$
real multiplications and $2 \times 2 + 4=8$ real additions, which can be denoted as
  a complexity of $3$ complex multiplications and $1$ complex addition.
   Accordingly,  the  complexity of (\ref{flopsOfGivens1928})
 can be deduced.

\section{The Conventional Givens Rotation Algorithm in \cite{MatrixComputationBook4th}
 Utilized to  Compute the Givens Matrix and the Corresponding Result}


The conventional complex
Givens rotation algorithm on \cite[p. 244]{MatrixComputationBook4th}
utilizes the real and imaginary parts of $u$ to
compute a  real Givens rotation with  $c_{\alpha}$, $s_{\alpha}$ and ${r}_{{u}}$,
and utilizes those parts of $v$ to
compute another  real Givens rotation with
${c}_{{\beta}}$,
$  {s}_{{\beta}}$ and ${r}_{{v}}$,
where
${r}_{{u}}$ and ${r}_{{v}}$ are defined by
\begin{subnumcases}{\label{uv2rurv2873}}
u=r_{u} e^{-i \alpha} &  \label{u2rue93988}\\
v=r_{v} e^{-i \beta}.  &  \label{v2rvebeta121}
\end{subnumcases}
Then  ${r}_{{u}}$
and ${r}_{{v}}$ are  utilized to compute  the third
real Givens rotation with $c_{\theta}$, $s_{\theta}$ and $r_{\theta}$, i.e.,
\begin{equation}\label{3rdGivens}
\left[\begin{array}{cc}{c_{\theta}} & {s_{\theta}} \\ {-s_{\theta}} & {c_{\theta}}\end{array}\right]^{T}\left[\begin{array}{l}{r_{u}} \\ {r_{v}}\end{array}\right]=\left[\begin{array}{l}{r_{\theta}} \\ {0}\end{array}\right],
 \end{equation}
 while $c_{\alpha}$, $s_{\alpha}$,
${c}_{{\beta}}$ and $  {s}_{{\beta}}$ are utilized to compute
\begin{equation}\label{eiphi2eibeta19281}
e^{i \phi}=e^{i(\beta-\alpha)}=\left(c_{\alpha} c_{\beta}+s_{\alpha} s_{\beta}\right)+i\left(c_{\alpha} s_{\beta}-c_{\beta} s_{\alpha}\right).
 \end{equation}
 Finally $c$ and $s$ in the Givens matrix
 $\left[\begin{array}{rr}{c} & {s} \\ {-s^*} & {c}\end{array}\right]$
 are obtained by
\begin{subnumcases}{\label{FastGivensCSCSCSuse}}
c = c_{\theta} &  \label{FastGivensCCCuse}\\
s = s_{\theta} {e^{i\phi }}.  &  \label{s2sThetaPhi322}
\end{subnumcases}

Now an efficient algorithm to compute the result ${r}$ in (\ref{givensForComplexROriginal})
is still required, which has not been given in \cite{MatrixComputationBook4th} .
 $e^{i \phi}=e^{i(\beta-\alpha)}$ in
(\ref{eiphi2eibeta19281})
can be substituted into (\ref{s2sThetaPhi322}) to obtain
\begin{equation}\label{s2sThetaPhi322New91281}
s = s_{\theta} e^{i(\beta-\alpha)}.
 \end{equation}
Then substitute
(\ref{s2sThetaPhi322New91281}), (\ref{FastGivensCCCuse}) and (\ref{uv2rurv2873}) into
%
(\ref{givensForComplexROriginal})
to obtain
\begin{equation}\label{givensForComplexR}
r=({c_{\theta}}{r_{u}}     -s_{\theta}{r_{v}})e^{-i \alpha}.
 \end{equation}
From (\ref{3rdGivens}), we can deduce
\begin{equation}\label{r2crusrv82932}
r_{\theta}={c_{\theta}}{r_{u}}     -s_{\theta}{r_{v}},
 \end{equation}
 and from (\ref{u2rue93988}),  we can deduce
 \begin{equation}\label{u2rue93988use32}
 e^{-i \alpha}=u/r_{u}.
 \end{equation}
Then (\ref{r2crusrv82932}) and
(\ref{u2rue93988use32})
 can be substituted into
 (\ref{givensForComplexR}) to obtain
 \begin{equation}\label{givensForComplexRsimpler}
r=(r_{\theta}/r_{u})u.
 \end{equation}

A real Givens rotation computed by equation (5.1.8) on \cite[p. 240]{MatrixComputationBook4th}
requires 4 real multiplications, 1 real additions, 1 real division and 1 real square root operation
~\footnote{When implementing equation (5.1.8)  in \cite{MatrixComputationBook4th} , one real division is reduced at the cost of adding two real multiplications, since
division and square root  are the most expensive real floating point operations ~\cite{Complex_Givens_Jul14b}.
}.
Then the conventional complex
Givens rotation on \cite[p. 244]{MatrixComputationBook4th} , which includes the above-described 3 real Givens rotations,
(\ref{eiphi2eibeta19281}),
(\ref{FastGivensCSCSCSuse}) and
 (\ref{givensForComplexRsimpler}),
 totally requires 20 real multiplications, 5 real additions, 4 real division and 3 real square root operation
 to compute $c$, $s$ and $r$ in (\ref{givensForComplexROriginal}).

\section{To explain the origin of (\ref{LengthModify}) that updates  the squared length of the rows in the inverse Cholesky factor}

 In our opinion,  (\ref{LengthModify}) was first implied by equation (7) in \cite{OriginalSqrtBlast}, which is the same
   as (\ref{equ9transform}).
Since a unitary transformation  will never change the length of any row vector in ${{\bf{F }}_{n}}$,
 it can easily be seen from (\ref{equ9transform}) that the squared length of the $j^{th}$
  row in ${{\bf{F }}_{n-1} }$ (i.e., ${\left| {{\bf{F }}_{n-1} }(j,:) \right|^2}$) can be computed by
 subtracting the squared absolute value of the $j^{th}$ entry of ${\bf{u}}_{n - 1}$ (i.e., $\left| {{\bf{u}}_{n - 1} }(j) \right|^2$)
 in the squared length of the $j^{th}$ row of
  ${{\bf{F }}_{n}}$ (i.e., $\left| {{\bf{F }}_{n}}(j,:) \right|^2$),
  which can be expressed by
    (\ref{LengthModify}).

Moreover,
we will show that
  (\ref{LengthModify}) can  be regarded as a specific instance of equation (A-34) on
\cite[p. 120]{PhdThesis}, i.e.,
\begin{equation}\label{Aphd34}
\mathbf{e}_{n-1}(i)=\mathbf{e}_{n}^{[-1]}(i)-\left|\mathbf{w}_{n-1}(i)\right|^{2} / \psi_{n},
 \end{equation}
 where $\mathbf{w}_{n-1}$ and $\psi_{n}$ are defined by equation (A-28) on
\cite[p. 118]{PhdThesis}, i.e.,
 \begin{equation}\label{Aphd28}
\mathbf{q}_{N-k}=\left[\begin{array}{ll}\mathbf{w}_{N-k-1}^{T} & \psi_{N-k}\end{array}\right]^{T},
 \end{equation}
 and are computed by equation (A-27) on \cite[p. 118]{PhdThesis},  i.e.,
 \begin{multline}\label{Aphd27}
\mathbf{q}_{N-k}^{H}=\overline{\mathbf{f}}_{(N-k) \times N}^{T} \mathbf{F}_{(N-k) \times N}^{H}
-   \\
\overline{\mathbf{w}}_{(N-k) \times k}^{T} \mathbf{C}_{k \times k} \mathbf{W}_{(N-k) \times k}^{H}.
 \end{multline}
Notice that the result of (\ref{Aphd34})
 is utilized in  line 10 on   \cite[p. 119]{PhdThesis}, to find   the undetected transmit signal with the highest signal-to-noise ratio (SNR).

 $\mathbf{e}_{N}(i)$ in
 (\ref{Aphd34}) is defined by equation (A-29) on \cite[p. 118]{PhdThesis}, i.e.,
$\mathbf{e}_{N}(i)=\mathbf{Q}_{N}(i, i)$,
where $\mathbf{Q}_{N}$ is defined by (A-15) on \cite[p. 116]{PhdThesis}, i.e.,
 $\mathbf{Q}_{N}=\mathbf{F}_{N} \mathbf{F}_{N}^{H}$. Then it can be seen that $\mathbf{e}_{N}(i)$ is the squared norm of the $i^{th}$  row in $\mathbf{F}_{N}$, i.e.,
  \begin{equation}\label{LengthModifyCompare29113}
\mathbf{e}_{N}(i) = \left| {{\bf{F }}_{N}}(i,:) \right|^2.
 \end{equation}

  In (\ref{Aphd27}),  the $k \times k$ matrix  $\mathbf{C}_{k \times k}$ and the $(N-k) \times k$ matrix $\mathbf{W}_{(N-k) \times k}$
  are defined by equations (A-22) and (A-21) on \cite[p. 116]{PhdThesis}, respectively,
  and $\overline{\mathbf{w}}_{(N-k) \times k}$ is the last row in $\mathbf{W}_{(N-k) \times k}$,
  as described in the first line under equation (A-27) on \cite[p. 118]{PhdThesis}. Then
  it can be seen that
    when $k=0$,
    $\overline{\mathbf{w}}_{(N-k) \times k}$,
   $\mathbf{C}_{k \times k}$ and  $\mathbf{W}_{(N-k) \times k}$ all become empty matrices,
   and
  (\ref{Aphd27})
   becomes
    \begin{equation}\label{q2fFMM23090ad}
\mathbf{q}_{N}^{H}=\overline{\mathbf{f}}_{N\times N}^{T}\mathbf{F}_{N\times N}^{H}.
 \end{equation}
 Notice that  $\mathbf{F}_{N\times N}$ in (\ref{q2fFMM23090ad}) satisfies
 \begin{equation}\label{}
\mathbf{F}_{N\times N}^{{}}=\mathbf{F}_{N}^{{}},
 \end{equation}
as can be seen from the definition of  $\mathbf{F}_{n\times N}$ in line 2 on \cite[p. 117]{PhdThesis}.

It is defined in the first line under (A-27) that  $\overline{\mathbf{f}}_{N\times N}^{T}$ is the last row in $\mathbf{F}_{N\times N}^{{}}$. Then
$\mathbf{F}_{N\times N}^{{}}=\mathbf{F}_{N}$  block triangularized by (\ref{equ9transform})
and the corresponding last row $\overline{\mathbf{f}}_{N\times N}^{T}$
can be substituted into
(\ref{q2fFMM23090ad})
 to obtain
${\bf{q}}_N^H = \left[ {\begin{array}{*{20}{c}}
{{\bf{0}}_{N - 1}^T}&{{\lambda _N}}
\end{array}} \right]\left[ {\begin{array}{*{20}{c}}
{{\bf{F}}_{N - 1}^H}&{{{\bf{0}}_{N - 1}}}\\
{{\bf{u}}_{N - 1}^H}&{{\lambda _N}}
\end{array}} \right]$, i.e.,
\begin{equation}\label{}
{\bf{q}}_N = \left[ {\begin{array}{*{20}{c}}
{{\lambda _N}{\bf{u}}_{N - 1}^H}&{\lambda _N^2}
\end{array}} \right]^H,
 \end{equation}
which is compared with
(\ref{Aphd28})
 to deduce
 \begin{subnumcases}{\label{wMlamdaupsid389ds}}
{{\bf{w}}_{N - 1}^{} = {\lambda _N}{\bf{u}}_{N - 1}^{}} &  \label{}\\
{{\psi _N} = \lambda _N^2}.  &  \label{}
\end{subnumcases}

Finally, let us substitute (\ref{wMlamdaupsid389ds}) into
 (\ref{Aphd34})
 to deduce
${{\mathbf{e}}_{N-1}}(i)=\mathbf{e}_{N}^{[-1]}(i)-{{\left| {{\lambda }_{N}}\mathbf{u}_{N-1}^{{}} \right|}^{2}}/\lambda _{N}^{2}$, i.e.,
\begin{equation}\label{e2eMu23898awe3d}
{{\mathbf{e}}_{N-1}}(i)=\mathbf{e}_{N}^{[-1]}(i)-{{\left| \mathbf{u}_{N-1}^{{}} \right|}^{2}},
 \end{equation}
 and then
substitute (\ref{LengthModifyCompare29113}) into (\ref{e2eMu23898awe3d}) to obtain  (\ref{LengthModify}),
to verify
 that (\ref{Aphd34}) can be equivalent to (\ref{LengthModify}).
 Accordingly,
 we have verified that
  (\ref{LengthModify}) can  be regarded as a specific instance of (\ref{Aphd34}), i.e.,
  equation (A-34) on
\cite[p. 120]{PhdThesis}.
\ifCLASSOPTIONcaptionsoff
  \newpage
\fi


\begin{thebibliography}{1}

\bibitem{zhfvtc2008} H. Zhu, W. Chen, B. Li, and F. Gao, ``An Improved Square-Root Algorithm for V-BLAST Based on
Efficient Inverse Cholesky Factorization", \emph{IEEE Trans. Wireless Commun.}, vol. 10,
no. 1, Jan. 2011.

\bibitem{BLASTtransSP2015} K. Pham and K. Lee,  ``Low-Complexity SIC Detection Algorithms for
 Multiple-Input Multiple-Output Systems", \emph{IEEE
Trans. on Signal Processing}, pp. 4625-4633, vol. 63, no. 17, Sept. 2015.

\bibitem{transSPcomments} H. Zhu and Y. Wu, ``Comments on `Low-Complexity SIC Detection
Algorithms for Multiple-Input Multiple-Output
Systems'", submitted to \emph{IEEE Trans. on Signal Processing}.




\bibitem{Complex_Givens_Jul14b}
D. Bindel, J. Demmel, W. Kahan and O. Marques, ``On Computing Givens
rotations reliably and efficiently", \emph{ACM Transactions on
Mathematical Software (TOMS) archive}, vol. 28 , Issue 2, June 2002.
Available online at: www.netlib.org/lapack/lawns/downloads/.




 \bibitem{OriginalSqrtBlast} B. Hassibi, ``An efficient square-root algorithm for BLAST",
\emph{Proc. IEEE Int. Conf. Acoustics, Speech, and Signal
Processing, (ICASSP '00)}, pp. 737-740, June 2000.

 \bibitem{PhdThesis} Hufei Zhu, ``Low Complexity Interference Cancellation Receivers for MIMO
Systems", \emph{Doctoral thesis, Shanghai Jiao Tong University, Shanghai, China}, Sept.  2013.
Available online at: http://wnt.sjtu.edu.cn/people/thesis/zhuhufei.pdf.

\bibitem{MatrixComputationBook}
 G. H. Golub and C. F. Van Loan, \emph{Matrix Computations}, Johns Hopkins University Press,
 Baltimore, MD, 3rd edition, 1996.

 \bibitem{MatrixComputationBook4th}
 G. H. Golub and C. F. Van Loan, \emph{Matrix Computations}, Johns Hopkins University Press,
 Baltimore, MD, 4th edition, 2013.


%
%
%
%
%
%

\end{thebibliography}
\end{document}